# MIIDL: a Python package for microbial biomarkers identification powered by interpretable deep learning

Jian Jiang

## Abstract

**Summary:** Detecting microbial biomarkers used to predict disease phenotypes and clinical outcomes is crucial for disease early-stage screening and diagnosis. Most methods for biomarker identification are linear-based, which is very limited as biological processes are rarely fully linear. The introduction of machine learning to this field tends to bring a promising solution. However, identifying microbial biomarkers in an interpretable, data-driven and robust manner remains challenging. We present MIIDL, a Python package for the identification of microbial biomarkers based on interpretable deep learning. MIIDL innovatively applies convolutional neural networks, a variety of interpretability algorithms and plenty of pre-processing methods to provide a one-stop and robust pipeline for microbial biomarkers identification from high-dimensional and sparse data sets.
**Availability:** Source code is available on GitHub (https://github.com/chunribu/miidl/) under the MIT license. MIIDL is operating system independent and can be installed directly *via* pip or conda.
**Contact:** chunribu@mail.sdu.edu.cn

## 1 Introduction

Microbial biomarkers play an important role in disease research (Dinan and Cryan, 2017). Detecting microbial biomarkers with high contribution to disease phenotypes is crucial for early-stage screening and diagnosis. Symptoms are not obvious to discover at early stages of certain diseases, and medical checkups can damage the body to varying extents. Fecal microbial sequencing is non-invasive and has no harm to the human body, which proved another choice for the diagnosis of certain diseases (Gilbert, et al., 2016).

The abundance of gut microbes is dynamic and heterogenetic, and identifying biomarkers from operating taxonomic units (OTUs) often encounters the "curse of dimensionality" as features are even more than observations. Many methods for biomarker identification are linear-based, which is very limited as biological processes are rarely linear. The emergence of machine learning seems to bring solutions to this problem (Libbrecht and Noble, 2015). There have been attempts and applications for biomarkers identification with random forests, support vector machines, and artificial neural networks, etc. Although high accuracies have been achieved, some shortcomings still exist in those schemes: 1) these models are usually "black boxes" which can barely provide biological and clinical interpretations; 2) prior feature selections are needed before modeling for high accuracy. ECMarker (Jin, et al., 2021) is an interpretable machine learning model predicting gene expression biomarkers which made a great demonstration in spite of sacrificing part of learning capacity and generalization ability for revealing underlying regulatory mechanisms.

Identifying microbial biomarkers in an interpretable, data-driven and robust manner remains challenging. To address this challenge, MIIDL (Markers Identification with Interpretable Deep Learning), a Python package for the identification of microbial biomarkers based on interpretable deep learning, is presented. MIIDL innovatively applies convolutional neural networks (CNNs) and a variety of interpretability algorithms, along with plenty of pre-processing methods, to provide a one-stop and robust pipeline for microbial biomarkers identification from high-dimensional and sparse data sets.

## 2 Materials and methods

As is illustrated in fig. 1, the first part of MIIDL workflow is data pre-processing, which is optional but vital in many cases. MIIDL requires one or two table(s) as input, with each row refers to an observation (sample), and each column a feature. A "Group" column will be auto-detected as labels in supervised learning.

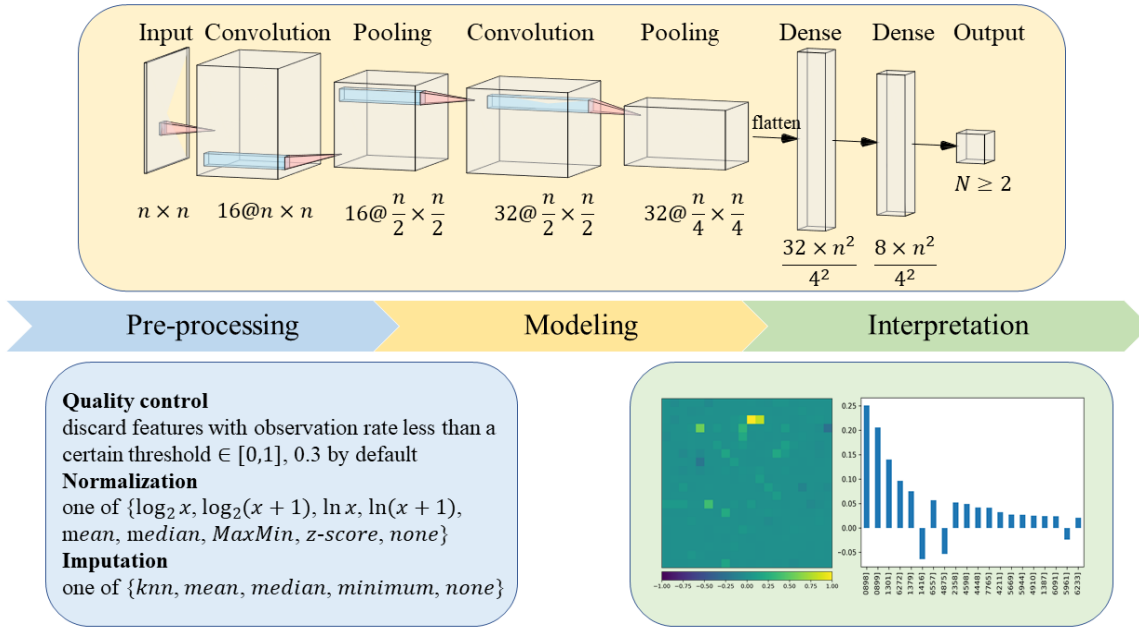

Fig. 1 The MIIDL workflow.

**Quality control.** Quality control is the first procedure. The observation rate is calculated for every feature and those less than a certain threshold is discarded. The threshold can be set to any value between 0 and 1.

**Normalization.** MIIDL offers many commonly used normalization methods to transform data and make samples more comparable, including `log₂x`, `log₂(x+1)`, `lnx`, `ln(x+1)`, `mean`, `median`, `maxmin`, `zscore`, and `none`. A `none` method in MIIDL is always used as a placeholder when you neither want to explicitly omit a step nor actually transform data.

**Imputation.** By default, this step is inactivated, as MIIDL is designed to solve problems including sparseness. But imputation can be useful in some cases. So, commonly used methods are available if needed, which include `knn`, `mean`, `median`, `minimum` and `none`.

**Reshaping.** The pre-processed data also need to be zero-completed to a certain length, so that a convolutional neural network (CNN) model can be applied. The conforming length can be calculated automatically following the rule that the number after the square root is divisible by 4, because every sample is a vector that will be reshaped to an n×n matrix and the downstream CNN model requires the width n can be divisible by 2 twice.

**Modeling.** A pre-designed CNN classifier is trained for discrimination. The model is adapted from the classical LeNet-5 (LeCun, et al., 2014) which is a simple convolutional neural network but powerful in various fields. Supervised learning for classification is performed and two or more classes (categories) are supported. PyTorch (Paszke, et al., 2019) serving as the backend is needed.

**Interpretation.** Captum (Kokhlikyan, et al., 2020) is dedicated to model interpretability for PyTorch, on which this step depends heavily. Integrated gradients (IG) algorithm is known as axiomatic attribution for deep networks, which is used to extract rules from a network and debug networks (Sundararajan, et al., 2017). In MIIDL, IG algorithm is recommended and is set as default, while multiple options are available including

`IntegratedGradients`, `Saliency`, `DeepLift`, `DeepLiftShap`, `InputXGradient`, `GuidedBackprop`, `GuidedGradCam`, `Deconvolution`, `FeaturePermutation`, `KernelShap` and LRP.

After the analysis is completed, there will be 3 files generated under the working path:

1) `FeatureImportance.tsv`: a tab-delimited text file containing importance, each column refers to a feature (species) and each row a sample;
2) `FeatureImportances.pdf`: importance values of features are plotted as a heatmap according to the order in `FeatureImportance.tsv` row by row;
3) `TOP20_KeyFeatures.pdf`: mean importance values are calculated and sorted. The 20 features with the largest absolute values are plotted in the bar chart.

## 3 Results

In this study, MIIDL is used to detect microbial biomarkers of colorectal cancer (CRC) and predict clinical outcomes. A dataset of fecal microbial mOTUs from 60 CRC patients and 60 healthy controls (labeled as CTR) is used, which is from a published study (Wirbel, et al., 2019). Specifically, the "German (DE) study" in the paper. The data set and the analysis process with detailed explanations are recorded and available on GitHub (https://github.com/chunribu/miidl/blob/main/Tutorials.ipynb). The trained classifier achieves an accuracy of 86.1 %. IG algorithm is applied to evaluate the importance of gut microbes to the CRC phenotype, potential microbial markers are obtained according to the corresponding importance values.

## 4 Implementation and availability

MIIDL is implemented in Python 3 and can be freely installed from Python Package Index (https://pypi.org/project/miidl) or Bioconda (https://anaconda.org/bioconda/miidl). Source code is available on GitHub (https://github.com/chunribu/miidl/ and https://chunribu.github.io/miidl/) under the MIT license.